\documentclass{PoS}

\title{Black Hole Lightning from the Peculiar Gamma-Ray Loud Active Galactic Nucleus\,IC 310}

\ShortTitle{Black Hole Lightning from IC\,310}

        \author{\speaker{Dorit Eisenacher Glawion}$^{a}$, Julian Sitarek$^{b,c}$, Karl Mannheim$^{a}$, Pierre Colin$^{d}$, for the MAGIC Collaboration, Matthias Kadler$^{a}$, Robert Schulz$^{e,a}$, Eduardo Ros$^{f,g,h}$, Uwe Bach$^{f}$, Felicia Krau{\ss}$^{e,a}$, J\"orn Wilms$^{e}$\\
        $^{a}$ Universit\"at W\"urzburg, D-97074 W\"urzburg, Germany\\
        $^{b}$ University of \L\'{o}d\'{z}, PL-90236 Lodz, Poland\\
        $^{c}$ IFAE, Campus UAB, E-08193 Bellaterra, Spain\\
        $^{d}$ Max-Planck-Institut f\"ur Physik, D-80805 M\"unchen, Germany\\
        $^{e}$ Dr. Remeis Sternwarte \& ECAP, Universit\"at Erlangen-N\"urnberg, ECAP, D-96049 Bamberg, Germany\\
        $^{f}$ Max-Planck-Institut f\"ur Radioastronomie, D-53121 Bonn, Germany,\\
        $^{g}$ Observatori Astron\`{o}mic, Universitat de Val\`{e}ncia, E-46980 Paterna, Val\`{e}ncia, Spain,\\
        $^{h}$ Departament d'Astronomia i Astrof\'{i}sica, Universitat de Val\`{e}ncia, E-46100 Burjassot, Val\`{e}ncia, Spain\\
        E-mail: \email{Dorit.Glawion@astro.uni-wuerzburg.de}}

\abstract{The nearby active galaxy IC\,310, located in the outskirts of the Perseus cluster of galaxies is a bright and variable multi-wavelength emitter from the radio regime up to very high gamma-ray energies above 100\,GeV. Originally, the nucleus of IC\,310 has been classified as a radio galaxy. However, studies of the multi-wavelength emission showed several properties similarly to those found from blazars as well as radio galaxies.
In late 2012, we have organized the first contemporaneous multi-wavelength campaign including radio, optical, X-ray and gamma-ray instruments. During this campaign an exceptionally bright flare of IC\,310 was detected with the MAGIC telescopes in November 2012 reaching an averaged flux level in the night of up to one Crab above 1\,TeV with a hard spectrum over two decades in energy. The intra-night light curve showed a series of strong outbursts with flux-doubling time scales as fast as a few minutes. The fast variability constrains the size of the gamma-ray emission regime to be smaller than 20\% of the gravitational radius of its central black hole. This challenges the shock acceleration models, commonly used to explain gamma-ray radiation from active galaxies. Here, we will present more details on the MAGIC data and discuss several possible alternative emission models.}

\FullConference{The 34th International Cosmic Ray Conference,\\
		30 July- 6 August, 2015\\
		The Hague, The Netherlands}

\begin{document}

\section{Introduction}

Relativistic jets are found in many astronomical systems, e.g., as part of the so-called active galactic nucleus (AGN) which hosts a supermassive black hole its center, surrounded by an accretion disk and the jets extending perpendicular to the disk. It is thought that particles in the jets can be accelerated to extreme energies and thus, contribute to the ultra-high energy cosmic rays with energies of $E>10^{18}$\,eV. Current key topics of the research of relativistic jets concern the connection between the black hole and the jet, the jet structure, and the jet formation process. Furthermore, the emission mechanisms at high energies are still a matter of debate. Hadronic and leptonic radiation processes leading to the observed gamma-ray emission are being discussed as well as the location of this emission region. Some of these topics can be studied by imaging the objects in different frequency ranges. However, even in the radio regime, where the best angular resolutions can be achieved, observations are yet capable of resolving the region close to the central engine where the jet formation process takes place. The best image of this region so far has been obtained for M87 by \cite{doelman} at 230\,GHz with a size of $11.0\pm0.8$ gravitational radii. \\
An alternative way to study these objects in this respect is to investigate the variability of the very-high-energy flux. The size of the emission region is constrained by the finite speed of light and causality. Hence, fast variability restricts the size of the region to be smaller than the light crossing time. In the gamma-ray energy range the imaging air Cherenkov telescopes provide a large collection area and hence, a possibility to observe flux variations on time scales of minutes. Such rapid events have been measured by H.E.S.S. and MAGIC from PKS\,2155$-$304 and Mrk\,501 in the last decade \cite{aharonian07, albert07}. These two objects are classified as blazars. According to the unified scheme, blazars are AGN for which the angle between the jet-axis and the line-of-sight is small ($\theta<10^\circ$). 
At these small angles, large Lorentz factors in the outflow will imply high Doppler factors, which have been related to high-energy emission \cite{begelman08}.\\     
\begin{figure}
    \centering
    \includegraphics[width=10cm]{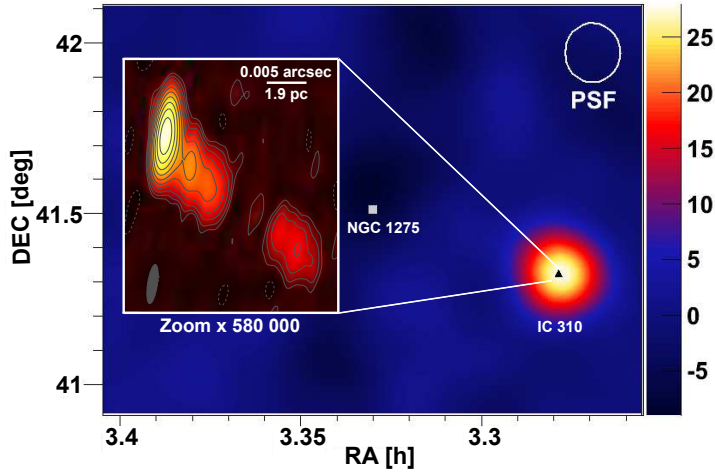}
    \caption{Combined image of the significance map measured by MAGIC above 300\,GeV on November 12-13, 2012 of the Perseus cluster of galaxies and the radio map obtained from EVN observations on October 29, 2012 at 5.0\,GHz. The figure is reprinted from \cite{aleksic14b}.}
    \label{skymapMAGICradioEVN}	
\end{figure}
The lenticular (S0, $z=0.0189$) galaxy IC\,310 hosts an AGN with a
supermassive black hole with a mass of $3\times10^8\,M_{\odot}$ inferred from the M-$\sigma$ relation (see \cite{aleksic14b}). This object has been detected at high energies above 30 GeV with \textit{Fermi}/LAT \cite{neronov10} as well as at very high energies above 260 GeV by MAGIC \cite{aleksic10, aleksic14a}. In the past, the
object was considered to be a head-tail radio galaxy with a large-scale ($\sim\,400$\,kpc projected) radio jet \cite{sijbring98}. However, Very Long Baseline Interferometry (VLBI) observations yielded that the jet shows the same direction in the sky from parsec to kiloparsec scales\cite{kadler12, aleksic14b}. The inner jet appears to be blazar-like with a missing counter jet, presumably due to relativistically boosted emission (see also Fig.~\ref{skymapMAGICradioEVN}, inset). From the total (de-projected) length of the kpc-scale jet and the VLBI observations, we can constrain the angle between the jet-axis and the line-of-sight to be $10\lesssim\theta\lesssim20$ making IC\,310 a borderline object between blazars and radio galaxies. Further indications for transitional behavior between a radio galaxy and a blazar was found in IC\,310 in various energy regimes as summarized in \cite{aleksic14a}.

\section{MAGIC observations and results}

As part of an extensive multi-wavelength campaign in late 2012 to early 2013, IC\,310 has been observed by the MAGIC telescopes. In the night of November 12-13, 2012 a bright flare was detected (see Fig.~\ref{skymapMAGICradioEVN}) with a significance of $\sim32\,\sigma$ above 300\,GeV \cite{aleksic14b}. 

\begin{figure}
    \centering
    \includegraphics[width=12cm]{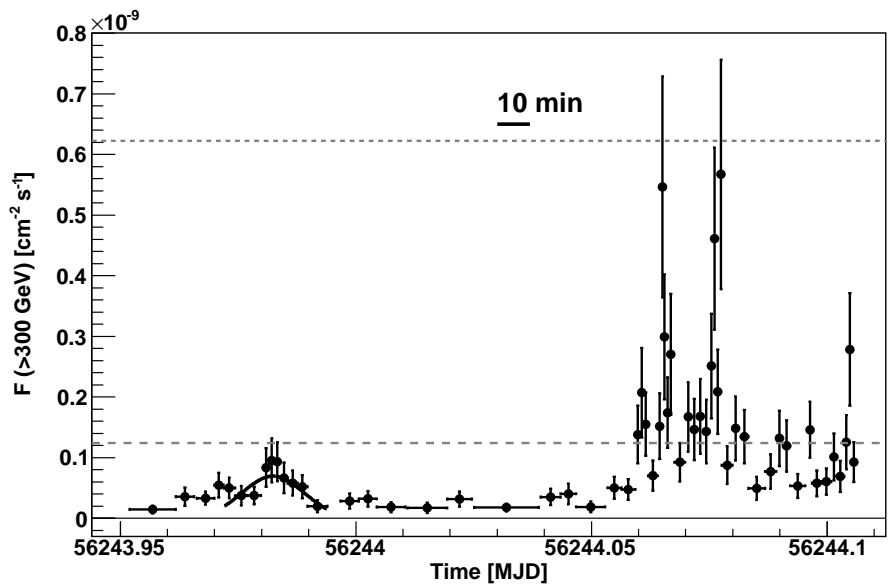}
    \caption{Light curve of the flare on November 12-13, 2012 measured by MAGIC above 300\,GeV. The pre-flare from 23:19 to 23:51 (MJD 56243.972--56243.994) is fitted with a Gaussian function with a standard deviation of $(9.5\pm1.9)$\,min. For comparison, the two dashed gray lines indicated the Crab Nebula flux and five times the flux of the Crab Nebula. The figure is reprinted from \cite{aleksic14b}.}
    \label{lightcurveMAGIC}	
\end{figure}

\begin{figure}
    \centering
    \includegraphics[width=8cm]{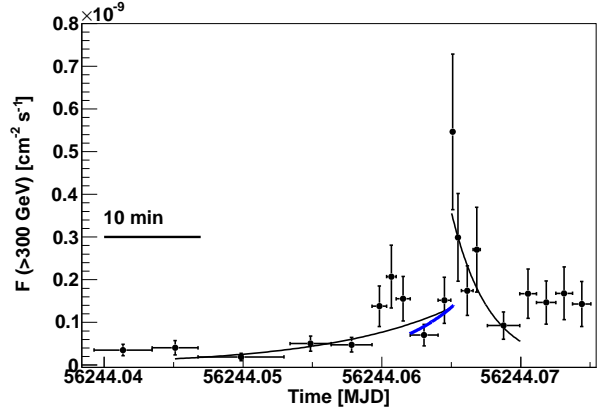}
    \caption{Zoom of the MAGIC light curve of the flare calculated above 300\,GeV into the time range 00:57 (MJD 56244.04) to 01:40 (MJD 56244.07). The black lines show exponential fits to the light curve to the rising and decay phases of the peak. The blue line shows the fit corresponding to the slowest doubling time necessary to explain the rising part of the flare at confidence level of 95\%. The figure is reprinted from \cite{aleksic14b}.}
    \label{lightcurveMAGIC_zoom}	
\end{figure}

During this night a mean flux of $(6.1\pm0.3)\times10^{-11}$\,cm$^{-2}$\,s$^{-1}$ above the same energy was measured which is four times higher than the high state flux measured during previous measurements in 2009/2010 \cite{aleksic14a}. Figure~\ref{lightcurveMAGIC} shows the light curve above 300\,GeV during the night. Fitting the light curve with a constant function reveals a $\chi^2/$d.o.f. of 199/58 corresponding to a probability of $2.6\times10^{-17}$. The light curve shows several sub-structures indicating fast variability on time scales of minutes. The pre-flare can be fitted with a Gaussian function with a standard deviation of $(9.5\pm1.9)$\,min.
Furthermore, we fitted the light curve with an exponential function in order to estimate the flux doubling time. For the sub-structure observed in the time range 00:57 (MJD 56244.04) to 01:40 (MJD 56244.07), a conservative value (95\,\% confidence level) for the flux doubling time of 4.88\,min could be found for the rising phase (Fig.~\ref{lightcurveMAGIC_zoom}). This value was obtained by fitting the rising part at 01:29--01:33 (MJD 56244.062--56244.0652) with many exponential functions, each time fixing the doubling time to a different value and calculating the probability of the fit. In the frame of reference of IC\,310, this doubling time corresponds to $4.9/(1+z)$\,min$=4.8$\,min. 

\begin{figure}
    \centering
    \includegraphics[width=9cm]{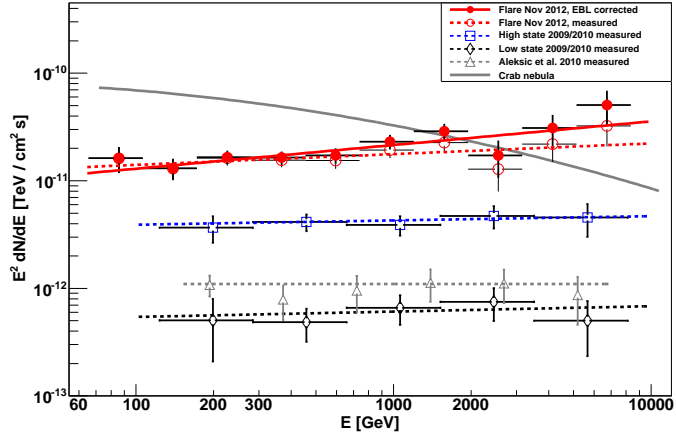}
    \caption{Averaged spectral energy distribution obtained by MAGIC during the flare shown together with previous measurements \cite{aleksic10, aleksic14a}. Open markers/dashed lines show the measured spectrum, whereas filled markes/solid lines indicate the spectrum corrected for the absorption due to the extragalactic background light according to the model by \cite{dominguez11}. For comparison, the spectral energy distribution from the Crab Nebula measured by MAGIC is shown in gray \cite{aleksic12}. The figure is reprinted from \cite{aleksic14b}.}
    \label{spectrumMAGIC}	
\end{figure}

The averaged spectrum of the night of the flare can be described with a simple power-law function $\mathrm{d}N/\mathrm{d}E=f_0\times (E/1\,\mathrm{TeV})^{-\Gamma}$ 
with a flux normalization of $f_0=(17.7\pm0.9_\mathrm{stat}\pm2.1_\mathrm{syst})\times10^{-12}$\,TeV$^{-1}$\,cm$^{-2}$\,s$^{-1}$ at 1\,TeV and a photon spectral index of $\Gamma=1.90\pm0.04_\mathrm{stat}\pm0.15_\mathrm{syst}$ between 70\,GeV and 8.3\,TeV. Due to the low redshift of IC\,310, the corrected spectrum due to the extragalactic background light changes by only 20\% in the flux normalization and $\sim0.1$ in the index. The spectral energy distribution from the flare and from previous measurements are shown in Fig.~\ref{spectrumMAGIC}. Compared to earlier observations, no significant spectral variability was observed. Our findings are similar to the ones in \cite{aleksic14a}. More details on the study of the spectral variability can be found in \cite{aleksic14b}. 
 
\section{Origin of the gamma-rays}

The standard model to explain of the production of gamma-rays in AGN is the so-called shock-in-jet model. In this scenario, particles are accelerated to high energies at shock fronts in the jet via Fermi acceleration and radiate, e.g., via the synchrotron-self Compton (SSC) process. Due to the magnetic field, the electron-positron plasma produces the observed synchrotron radiation at lower frequencies and in case of the SSC model inverse-Compton scattering on the same synchrotron photon field leads to the observed high-energy emission. The smallest scale of the shock front is determined by the smallest diameter of the jet, i.e., the jet base which is itself assigned by the event horizon of the black hole. In case of variability observed from PKS\,2155$-$304 and Mrk\,501, the radius of a spherical emission region is smaller than the event horizon $R_\mathrm{G}$. However, these observations are still consistent with the standard model due to the relativistic aberration and a large Doppler factor \cite{begelman08}. 

In case of IC\,310, the variability of $4.8$\,min constrains the radius of the emission region to be $\sim$\,20\% of the event horizon for the aforementioned mass of the black hole. Due to a rather large viewing angle between $10^\circ$ and $20^\circ$ and low Doppler factors observed with radio telescopes \cite{kadler12}, Doppler boosting can hardly solve this problem. Therefore, this fast variability cannot be easily explained with the standard shock-in-jet model. Furthermore, in such a small emission region as inferred from the fast variability, $\gamma\gamma$ pair production would lead to the absorption of the TeV gamma-rays making the emission impossible to be observed. The optical depth for 10\,TeV photons for this process can be calculated to be $\tau_{\gamma\gamma}(10\,\mathrm{TeV})\sim300(\delta/4)^{-6}(\tau_{\mathrm{var}}/1\,\mathrm{min})^{-1}(L_\mathrm{syn}/10^{42}\,\mathrm{erg}\,\mathrm{s}^{-1})$ assuming a Doppler factor of $\delta=4$ \cite{kadler12}, a variability time scale of $\tau_\mathrm{var}=1$\,min, and a synchrotron luminosity of $L_\mathrm{syn}=10^{42}\,\mathrm{erg}\,\mathrm{s}^{-1}$.  

\begin{figure}
    \centering
    \includegraphics[width=6cm]{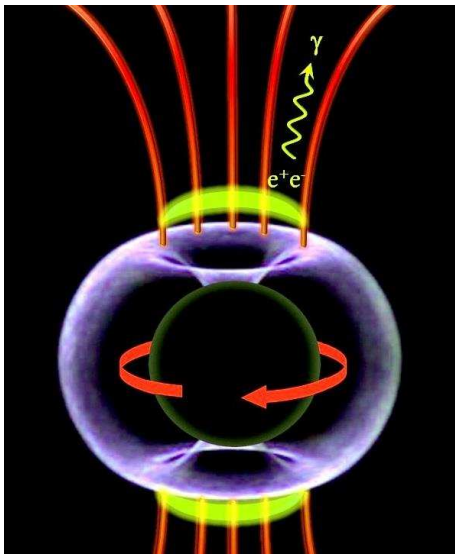}
    \caption{Magnetospheric model for the origin of the gamma-rays around a maximally rotation black hole with an event horizon of $R_\mathrm{G}$ (black sphere). The blue shape indicates the ergosphere of the black hole in which Poynting flux is generated by the frame-dragging effect. Due to the rotation, a charge-separated magnetosphere (red) is induced and vacuum gaps (yellow) are formed at the polar regions. Insight these gaps, particles can be accelerated to ultra-relativistic energies due to an electric field parallel to the magnetic field. The observed gamma-rays can be explained by inverse-Compton scattering and copious pair production via low-energy thermal photons originating from the accreted plasma. The figure is reprinted from \cite{aleksic14b}.}
    \label{gap}	
\end{figure}

We investigated several models that could explain the sub-horizon variability and anisotropic radiation (to avoid pair absorption), e.g., mini-jet structures within the jets \cite{giannios10}, jet-cloud/star interactions \cite{bednarek97, barkov12}, and magnetospheric models \cite{levinson11}. As extensively demonstrated in \cite{aleksic14b}, only the last model was able to explain all observed properties. Magnetospheric models are originally used to explain the high-energy emission from pulsars. Instead of a pulsar, a particle-starved magnetosphere is anchored to the ergosphere of the black hole which is assumed to spin maximally. Electrons and positrons can be accelerated in vacuum gaps due to electric fields that are parallel to magnetic fields. Here, the density of the charge carriers needs to be lower than the Goldreich-Julian charge density, otherwise a short-circuit occurs. Because the charge carrier density is proportional to the accretion rate of the system, the accretion rate needs to be low. Such a low rate can be assumed for IC\,310 as the jet power is rather low \cite{aleksic14b}. The fast gamma-ray variability observed from IC\,310 can be explained by a small gap height. This height is calculated to be $h\sim0.2\,R_{\mathrm{G}}$, assuming an accretion rate of $\dot{m}\sim10^{-4}$ in units of the Eddington rate and a maximally rotating black hole. The gaps can be located at the pole of the magentosphere and can point at different angles with the jet axis, e.g., at angles of $10^\circ-20^\circ$. The accelerated particles produce an electromagnetic cascade leading to a high multiplicity of the electron-positron pairs. After some time, the gap short-circuits and the particles move away with the jet. Then, the gap may reopen and produce new flares. The emission is produced by either inverse-Compton scattering on a background photon field, or by curvature radiation. According to \cite{mannheim93}, such cascading lead to a rather stable energy spectrum with a power-law index of 1.9 as observed from IC\,310.       

\section{Conclusions}

For first time, variability shorter than the event horizon light crossing time has been observed that is not conform with a simple shock-in-jet model. Instead, it is consistent with a magentospheric model for high-energy emission in AGN. To finally answer which kind of mechanism is responsible for ultra-rapid flux variability events, more observations are needed, e.g., with higher sensitivity as provided by the Cherenkov Telescope Array as this potentially allows to measure even faster flares as well as rapid flux variations from other, e.g., fainter AGN. Furthermore observing such events with simultaneous multi-wavelength coverage can help to constrain emission models.

\section{Acknowledgments}

We would like to thank
the Instituto de Astrof\'{\i}sica de Canarias
for the excellent working conditions
at the Observatorio del Roque de los Muchachos in La Palma.
The financial support of the German BMBF and MPG,
the Italian INFN and INAF,
the Swiss National Fund SNF,
the ERDF under the Spanish MINECO (FPA2012-39502), and
the Japanese JSPS and MEXT
is gratefully acknowledged.
This work was also supported
by the Centro de Excelencia Severo Ochoa SEV-2012-0234, CPAN CSD2007-00042, and MultiDark CSD2009-00064 projects of the Spanish Consolider-Ingenio 2010 programme,
by grant 268740 of the Academy of Finland,
by the Croatian Science Foundation (HrZZ) Project 09/176 and the University of Rijeka Project 13.12.1.3.02,
by the DFG Collaborative Research Centers SFB823/C4 and SFB876/C3,
and by the Polish MNiSzW grant 745/N-HESS-MAGIC/2010/0.
We thank also the support by DFG WI 1860/10-1. 
J. S. was supported by ERDF, the Spanish MINECO through FPA2012-39502, JCI-2011-10019 grants, and by Fundacja U\L. 
E. R. was partially supported by the
Spanish MINECO projects AYA2009-13036-C02-02
and AYA2012-38491-C02-01 and by the Generalitat
Valenciana project PROMETEO/2009/104, as well as
by the COST MP0905 action 'Black Holes in a Violent
Universe'. The European VLBI Network is a joint facility of European, Chinese, South African and other
radio astronomy institutes funded by their national research councils. The research leading to these results
has received funding from the European Commission
Seventh Framework Programme (FP/2007-2013) under grant agreement No. 283393 (RadioNet3).

\end{document}